\documentclass[journal=jpccck,manuscript=article]{achemso}
\usepackage[version=3]{mhchem} 
\usepackage{amsmath,amssymb}
\usepackage{mathrsfs}
\usepackage{graphicx}

\usepackage{xcolor}
\author{Masaru Oda}

\email{odamasa@mns.kyutech.ac.jp}
\phone{+81 93 884 3420}
\affiliation[Kyushu Institute of Technology]
{Department of Basic Sciences, Faculty of Engineering, Kyushu Institute of Technology, 1-1 Sensui-cho, Tobata-ku, Kitakyushu, Fukuoka, 804-8550 Japan}

\author{Kazuaki Yamato}
\affiliation[Kyushu Institute of Technology 2]
{Department of Electrical and Electronic Engineering, Graduate School of Engineering, Kyushu Institute of Technology, 1-1 Sensui-cho, Tobata-ku, Kitakyushu, Fukuoka, 804-8550 Japan}

\author{Jyunya Egashira}
\affiliation[Kyushu Institute of Technology 2]
{Department of Electrical and Electronic Engineering, Graduate School of Engineering, Kyushu Institute of Technology, 1-1 Sensui-cho, Tobata-ku, Kitakyushu, Fukuoka, 804-8550 Japan}

\author{Hisao Kondo}
\affiliation[Ehime University]
{Mathematics, Physics, and Earth Sciences, Graduate School of Science and Engineering, Ehime University, 2-5, Bunkyo-cho, Matsuyama, Ehime 790-8577 Japan}

\title[An \textsf{achemso} demo]
 {Room-temperature strong coupling of hexane-dispersed colloidal CdSe nanoplatelets in a microcavity composed of two Bragg reflectors}
\abbreviations{NPLs, RT, DBR}
\begin{document}
%
%
%
%
%
\begin{abstract}
 CdSe nanoplatelets (NPLs) are suitable for exploring strong light-matter coupling in semiconductor nanocrystal systems due to their giant oscillator strength and large exciton binding energy. Herein, we report on the facile fabrication and optical characterization of a $\lambda$/2 planar microcavity, which consists of two distributed Bragg reflectors with a hexane layer containing concentrated colloidal CdSe NPLs. Using a hexane solution layer instead of the typically used dried active layers makes the layer thin and flat, even under dense NPL conditions, without stressing or charging of the NPLs' surfaces. Reflectance spectra showed that strong light-matter coupling can be realized at room temperature and that the vacuum Rabi splitting energy is 53.5 meV. Intense photoluminescence (PL) emerges at the lower polariton branch where 25.1 meV (= $E$$\rm_{LO}$: longitudinal optical (LO)-phonon energy) below the energy of the polariton dark states, indicating that the relaxation from the dark states occurs efficiently in this microcavity owing to LO-phonon-assisted relaxation. We describe the reflectance and PL properties using the model that a cavity photon couples to a one-exciton state delocalized over nonuniformly orientated NPLs. This model contributes to an intuitive and quantitative understanding of the microcavity containing colloidal NPLs. 
\end{abstract}
\section{1. Introduction}
Cavity polaritons are quasi-particles that form through the strong coupling between cavity photons and excitons in semiconductor materials~\cite{Weisbuch_1992}. The formation leads to new optical and electronic properties, including large Rabi-splitting~\cite{Lidzey_1998} and Bose-Einstein condensation~\cite{BEC_Deng_2010} of the polaritons. Planar microcavities, consisting of a semiconductor active layer between two parallel mirrors whose distance is tuned to resonate the energies of the cavity photon with the exciton, are the simplest and most commonly used structures in past studies~\cite{Weisbuch_1992, polariton_chemistry, Lidzey_1998, BEC_Deng_2010}; they have been researched extensively for decades due to interest in their fundamental properties and prospective applications, such as extremely low-threshold lasers~\cite{Deng_2003} and efficient light-emitting devices~\cite{Stranius_2018}.
Many applications require room-temperature (RT) operation. However, at RT, polaritons form only in planar microcavities containing wide-gap semiconductors, i.e., ultraviolet-emitting semiconductors, such as GaN~\cite{GaN_2007}, ZnO~\cite{ZnO_Guillet_2011}, CuBr~\cite{CuBr_2012}, and perovskite CsPbCl$_2$~\cite{Su_3982}, when using inorganic semiconductors with a normal shape, i.e., a plate with the thickness of the same or near the distance between two mirrors. This limitation occurs because polariton formation at RT requires both intrinsic exciton stability and large oscillator strength of the exciton in the semiconductors~\cite{GUILLET_2016}.
Polaritons can form at RT in microcavities containing visible-emitting semiconductors when using nanocrystals (NCs), such as colloidal CdSe nanoplatelets (NPLs)~\cite{AgNP_2016, Yang_2022}, quantum dots (QDs)~\cite{prism_2011, CdSeQD_2021}, and WS$_2$~\cite{Flatten_2016}, WSe$_2$, and MoSe$_2$~\cite{Gillard_2021} monolayered plates. The advantage of NCs is due to their exciton confinement effects; for example, the excitons are stable at RT in CdSe NPLs whose thicknesses are several monolayers (MLs). This is attributed to the large exciton binding energy of 195-315 meV~\cite{Shornikova_2021}, in contrast to that in bulk crystals of 15 meV~\cite{Voigt_1979}. The large binding energy results from exciton confinement in extremely thin NPLs, in which there is a significant difference in the dielectric constants of the inner and outer parts of the crystals~\cite{Shornikova_2021}. Furthermore, the large binding energy strengthens the oscillator strength of the exciton. Additionally, the giant oscillator strength effect works on the excitons in the NPLs, owing to the in-plane coherent movement of the excitons~\cite{Naeem_2015}. According to these effects, the requirements for RT polariton formation are satisfied in NPLs. 
A recent study first reported on RT polariton lasing in a microcavity containing colloidal NC systems~\cite{Yang_2022}; the microcavity was composed of two distributed Bragg reflector (DBR) mirrors with a 1-µm-thick active layer containing CdSe NPLs. A typical minimum standard of planar microcavities' quality factor (Q) value for RT polariton lasing is generally considered to be $\sim$1000~\cite{GUILLET_2016} in inorganic semiconductor microcavities. Although the Q-value was only 115 of the NPL microcavity, the lasing was achieved with an ultralow threshold (0.5 $\mu$J cm$^{-2}$). The as-described lasing is mainly due to the large absorption and low-loss efficiencies of the polaritons in the microcavity, which are attributable to the advantages of the colloidal NPLs, specifically, a large absorption cross-section and a high quantum yield of photoluminescence (PL). In addition, the NPLs offer the advantage that their PL wavelength can be tuned over a wide range by varying their thickness or lateral-size control depending on the application. As such, there is great interest in NPLs as a promising candidate material for inorganic light-emitting polariton devices, especially in the visible regime. 
However, there have been few RT polariton studies of NPL microcavities to date~\cite{AgNP_2016, Yang_2022}, even when including the microcavities containing other colloidal NCs~\cite{prism_2011, CdSeQD_2021}. Besides, most of the NC microcavities are composed of single-layered mirrors, such as metallic ~\cite{AgNP_2016} or total reflection~\cite{prism_2011} mirrors, instead of the multi-layered DBR mirrors, which are widely used for lasing in planar microcavities due to their high reflectivity. One of the main reasons for the limitations is the difficulty in fabricating the microcavities. The active layer containing NPLs is usually made from a dried film consisting of the NCs and ligand organic molecules~\cite{AgNP_2016, Yang_2022}, with one exception~\cite{CdSeQD_2021}. To function as the planar microcavity, the dried film must be formed between two precisely parallel mirrors with a distance of adequate cavity thickness. Also, the film must contain a large amount of NPLs to induce a strong light-matter interaction. It is difficult to meet all requirements simultaneously, especially when the manufacturing includes the multi-layered DBR mirrors. Furthermore, the NCs embedded in dried film sometimes show an unexpected tail or band in the PL spectra due to effects related to NC surfaces, such as charging~\cite{oda_2010}, stressing, and aggregation.
This paper describes a facile fabrication method and optical characterization of a $\lambda$/2 microcavity with a 150-nm-thick hexane layer containing concentrated colloidal 4-ML-thick CdSe NPLs between two as-prepared DBR mirrors. The thin and flat layer can be formed using the solution layer, even under dense NPL conditions, without stressing or charging the NPLs' surfaces. We demonstrate RT polariton formation and report on a new finding of PL properties of the lower polariton related to longitudinal optical (LO)-phonon-assisted relaxation from the dark states to the lower polariton state. We describe the experimental results based on the model that the cavity photon couples to the one-exciton state delocalized over nonuniformly oriented NPLs; this is a partially extended model of the Tavis-Cummings model~\cite{TC1,TC2}.
\section{2. Results and discussion}
\subsection{2.1. Samples and methods}
\begin{figure}[htb]
 \centering
 \includegraphics[width=4in]
      {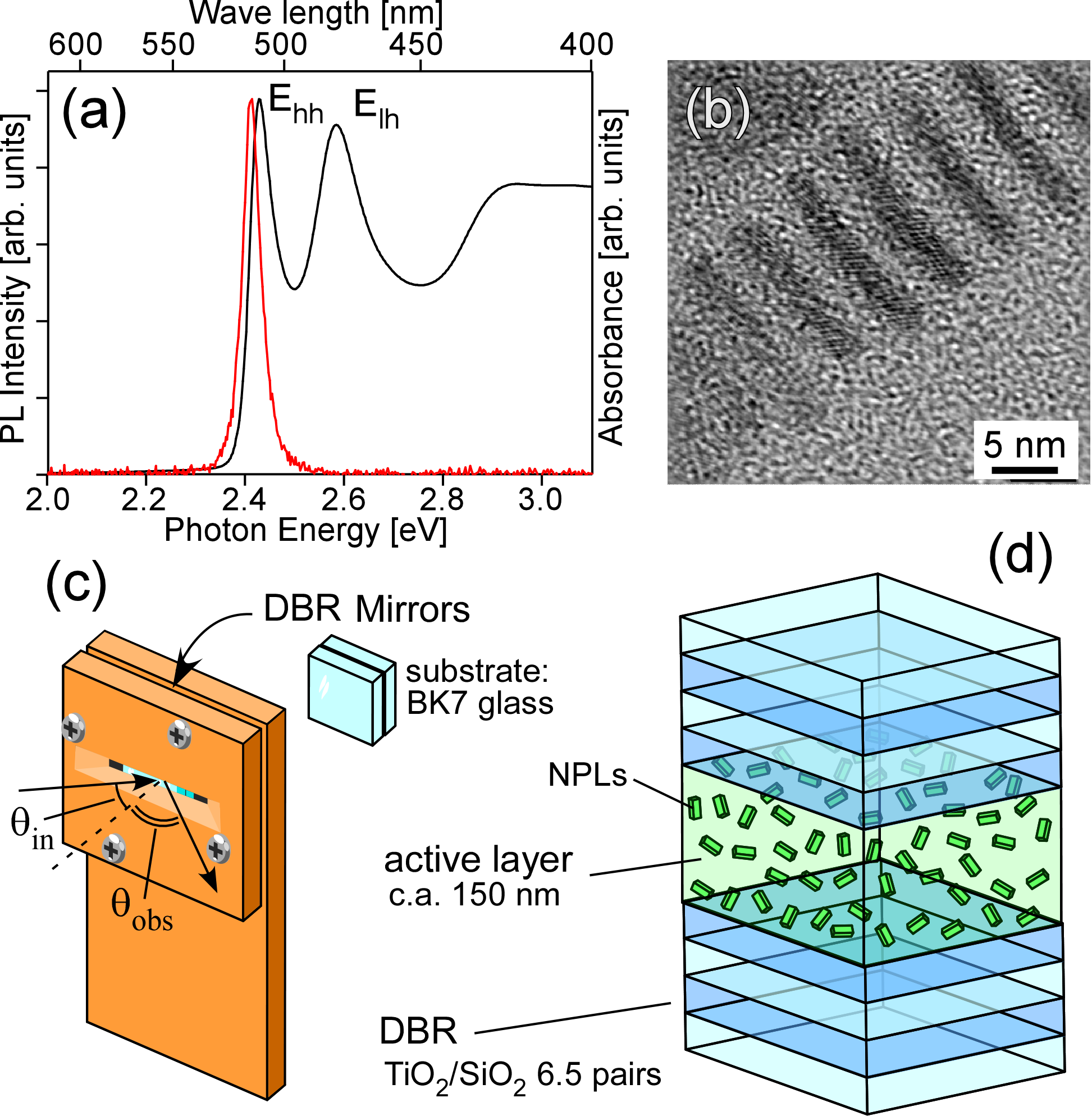}
 \caption{(a) Absorption (black) and photoluminescence (PL) (red) spectra of CdSe nanoplatelets (NPLs) in hexane solution at room temperature (RT). (b) Transmission electron microscopy image of the NPLs taken with a JEM-F200 system (JEOL Ltd., Tokyo, Japan). Several self-aligned NPLs are visible. Schematic diagrams of (c) the sample holder and (d) the microcavity, consisting of an NPL-dispersed solution layer and bottom and top TiO$_2$/SiO$_2$ DBR layers.}
 \label{sample}
\end{figure}

Oleic-acid-capped NPLs were synthesized using the method described by Bertrand et al. for the synthesis of zincblende NPLs~\cite{Bertrand_2016} (see Supporting Information (SI) A1). The black line in Fig.~\ref{sample}(a) shows the absorption spectrum of the CdSe NPLs in hexane at RT. The first and second absorption bands whose energies are $E_{\text{hh}}$ $\simeq$ 2.43 and $E_{\text{lh}}$ $\simeq$ 2.54 eV correspond to heavy-hole (hh) and light-hole (lh) exciton transitions, respectively. The $E_{\text{hh}}$ and $E_{\text{lh}}$ depend mainly on the NPLs' thickness, as the excitons in the NPLs are confined strongly in the [001] thickness direction; the energy values indicate that the NPLs' thickness is 4 MLs (1.4 nm)~\cite{Bertrand_2016}. The NPL solution is a mixture containing two components showing different shapes at a ratio of 4 to 1: the main component is a rectangle with an average length and width of 13.5 $\pm$ 0.5 nm and 2.5 $\pm$ 0.1 nm, respectively (Fig.~\ref{sample}(b)), and the other is roughly square (see SI A2). The absorption band of the hh-exciton is strong and narrow, with a full width at half maximum (FWHM) of 47 meV, which is attributed to the large oscillator strength of the confined excitons in CdSe NPLs with a precisely uniform thickness~\cite{Naeem_2015}. The red line shows the PL spectrum of hexane-dispersed NPLs under 400-nm (3.1 eV) illumination. The PL band, located at $E_\text{PL}$ $\simeq$ 2.41 eV, is due to radiative recombination of the hh-excitons.
In the following manner, we fabricated a planar microcavity composed of two DBR mirrors sandwiching an NPL-hexane solution layer. First, the prepared NPL-hexane solution was concentrated by heating near the boiling point of hexane under atmospheric pressure (see SI A3). After cooling, 5 $\rm{\mu}$L of the concentrated solution was dropped onto a custom-made DBR mirror purchased from SIGMA KOKI Ltd. (Tokyo, Japan) that consisted of 6.5-pair TiO$_2$/SiO$_2$ layers, with a designed wavelength of 518 nm, deposited on a 1-mm-thick substrate of BK7 glass. After the droplet was covered with another of the same structured mirror, we pushed out the excess solution by pushing near the central part of the upper mirror with plastic tweezers until its interference pattern changed from closely spaced lines to a ring (or two-centric rings). The ring pattern indicates that the thickness of the solution layer at the central part is the thinnest and near $\lambda$/2. After the sample was set in a copper holder, as shown in Fig.~\ref{sample}(c), the thickness and flatness of the solution layer were manually tuned by adjusting screws, confirming the interference pattern visually. Finally, the active layer was sealed with polyvinyl alcohol (PVA) by painting and drying a PVA aqueous solution on the side of the microcavity. The microcavity structure after the tuning is illustrated in Fig.~\ref{sample}(d). We consider the orientation direction of individual NPLs in the microcavity to be random, as they are dispersed throughout the hexane solution. Notably, the NPLs in Fig.~\ref{sample}(b) were self-aligned using a drying process, as usually occurs when preparing the sample for transmission electron microscopy observations.
We measured the angle-dependent reflection and PL spectra at RT for transverse magnetic (TM) polarized light to investigate the microcavity's mode structure and PL properties. In the reflection measurement, white light emitted from a tungsten-halogen lamp (SLS201L, Thorlabs, Newton, NJ, USA) illuminated the sample at incident angle $\theta_{\rm{in}}$, and the reflected light was observed with a viewing angle of $\theta_{\rm{obs}}$ (= $\theta_{\rm{in}}$), as illustrated in Fig. 1(c). In the PL measurement, 407.5-nm light ($E_\mathrm{exc} = $3.04 eV) from a pulsed diode laser (LDH-PC405, PicoQuant, Berlin, Germany) illuminated the sample at normal incidence ($\theta_{\rm{in}}$ = 0). The PL from the NPLs was observed with a viewing angle of $\theta_{\rm{obs}}$. The reflection and PL spectra were measured using a monochromator (SpectraPro-300i, Acton Research, Acton, MA, USA) with a charge-coupled device (CCD) detector (LN/CCD, Princeton Instruments, Trenton, NJ, USA) cooled by liquid nitrogen. The spot diameters of the illumination and the observation areas were approximately 0.6 mm. The light powers of both the tungsten ramp and the pulsed laser were set to be sufficiently weak.
\subsection{2.2. Model of light-matter strong coupling in CdSe NPLs}
\subsubsection{2.2.1. Strong coupling between photons and $N$ molecules}
\begin{figure}[htb]
 \centering
 \includegraphics[width=5.5in]
      {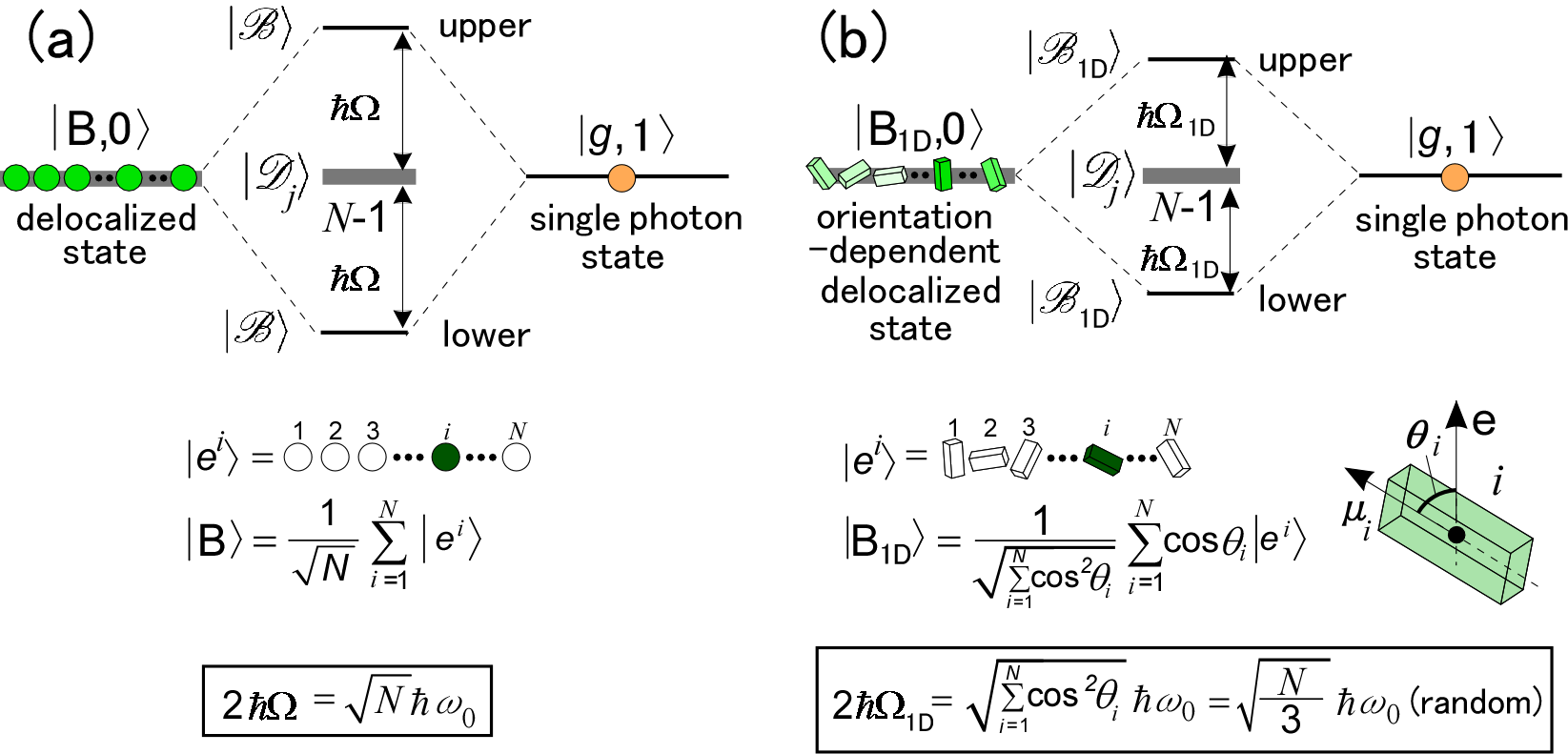}
  \caption{Schematic diagrams of the strong light-matter coupling between the single photon and the delocalized states over $N$ molecules (a) without and (b) with considering the orientation dispersion of the 1D transition dipole moment of the molecules. $\left| {e^i}\right\rangle$ is the state where the $i^{th}$ molecule is in the excited state, and the others are in their ground states.}
 \label{model}
\end{figure}

 We consider the light-matter strong coupling between photons in a single-mode and $N$ molecules in the microcavity on the basis of the TC model~\cite{TC1,TC2}. The Hamiltonian of $N$ molecules with two energy levels in a lossless microcavity is described by
\begin{eqnarray}
H = {E_\text{ph}}{\hat a^\dag }\hat a + \sum\limits_{i = 1}^N {E_\text{ex}^i\hat \sigma _ + ^i\hat \sigma _ - ^i} - \sum\limits_{i = 1}^N {\hbar {\omega_i}}\left( {{{\hat a}^\dag }\hat \sigma _ - ^i + \hat a\hat \sigma _ + ^i} \right) 
 \label{tc_H}
\end{eqnarray}
on the assumption of ignoring the variation of the electric field amplitude in the microcavity. $E_\text{ph}$ and $E_\text{ex}^i$ are the cavity photon energy and the exciton energy in the $i^{th}$ molecule, respectively. $\hat{a}^\dag$ ($\hat{a}$) is the cavity photon creation (annihilation) operator, and $\hat{\sigma} _ + ^i $ ($\hat{\sigma} _ - ^i $) is the exciton creation (annihilation) operators in the $i^{th}$ molecule, and $\hbar \omega_i $ is the strength of the light-matter interaction of the $i^{th}$ molecule. The lowest excitation states in the system, of which the total excitation number $N_{exc}$ $(\equiv {\hat a^\dag }\hat a + \sum\limits_{i = 1}^N {\hat \sigma _ + ^i\hat \sigma _ - ^i})$ = 1, can be described by linear combinations of $N$+1 basis: $\left| {g,1} \right\rangle$ and $\left| {{e^i},0} \right\rangle$ ($i = 1,2, \cdot \cdot \cdot ,N$). The $\left| {g,1} \right\rangle$ is the single photon state where all molecules are in their ground states, and simultaneously, a single cavity photon exists. The $\left| {{e^i},0} \right\rangle$ are the single exciton states where the $i^{th}$ molecule is excited; the others are in their ground states, and simultaneously, no cavity photon exists. 
When the $N$ molecules are identical, i.e., $E_\text{ex}^i$ = $E_\text{ex}$ and $\hbar \omega_i$ = $\hbar \omega_0$, the lowest excitation states can be expressed by two light-matter hybrid bright states $\left|\mathscr{B} \right\rangle$, referred to as lower and upper polaritons, and ($N$-1)-fold degenerate dark states $\left| \mathscr{D}_j \right\rangle$
($j = 1,2, \cdot \cdot \cdot ,N-1$) with the energy of $E_\text{ex}$~\cite{polariton_chemistry, Fink_2009}, as illustrated in Fig.~\ref{model}(a).
The $\left|\mathscr{B} \right\rangle$ is defined as
\begin{eqnarray}
\left|\mathscr{B} \right\rangle \equiv \alpha \left| g,1\right\rangle+\beta\left| \rm{B},0 \right\rangle,
 \label{bright states}
\end{eqnarray}
where $\alpha$ and $\beta$ are the coefficients satisfying $ \left| \alpha \right|^{2}$+$ \left| \beta \right|^{2}$=1. $\left| \rm{B,0}\right\rangle$
is the delocalized exciton state where the $N$ molecules are in the delocalized permutationally invariant exciton state $\left| \rm{B}\right\rangle$
($\equiv$ $N$$^{-\frac{1}{2}}$$\sum\limits_{i = 1}^N$$\left| {{e^i}} \right\rangle$ ), and simultaneously, no cavity photon exists.
The bright states $\left|\mathscr{B} \right\rangle$ are formed by the light-matter strong coupling between the delocalized exciton state $\left| \rm{B,0}\right\rangle$ and the single photon state $\left| g,1\right\rangle$. The dark states $\left| \mathscr{D}_j \right\rangle$, where the $N$ molecules are in delocalized non-totally symmetric states, show no coupling to the single photon state, due to the states' symmetry. The stationary Schr\"{o}dinger equation for the bright states $\left|\mathscr{B}\right\rangle$ can be simplified as
\begin{eqnarray}
\left( {\begin{array}{*{20}{c}}
{{E_\text{ph}}}&{\hbar \Omega }\\
{\hbar \Omega }&{{E_\text{ex}}}
\end{array}} \right)\left( {\begin{array}{*{20}{c}}
\alpha \\
\beta 
\end{array}} \right) = E_{u,l}\left( {\begin{array}{*{20}{c}}
\alpha \\
\beta 
\end{array}} \right),
 \label{matrix_2}
\end{eqnarray}
where $\hbar \Omega$ (= $\sqrt N \hbar \omega_0$) is the interaction energy between $\left| \rm{B},0 \right\rangle$ and $\left| g,1\right\rangle$. $E_{u,l}$ are the energies of the upper and lower polaritons described by
\begin{eqnarray}
E_{u,l} = \frac{{{E_\text{ph}} + {E_\text{ex}}}}{2} \pm \frac{1}{2}\sqrt {{{({E_\text{ph}} - {E_\text{ex}})}^2} + {{(2\hbar \Omega )}^2}} .
 \label{polariton energies}
\end{eqnarray}
Note that when $N$=1 is substituted into the Eqs.~(\ref{matrix_2}) and~(\ref{polariton energies}), the equations are the same as those in the so-called coupled oscillator model~\cite{Skolnick_1998} that describes the strong coupling between a cavity photon and a single molecule with a two-level system.
\subsubsection{2.2.2. Strong coupling between a photon and $N$ molecules having an anisotropic dipole moment}
In the case that the $N$ molecules have a 1D transition dipole moment $\boldsymbol{\mu}$$_{i}$, and a photon is polarized with the oscillation direction along the unit vector $\textit{\textbf{e}}$, the strength of the light-matter interaction of the $i^{th}$ molecule $\hbar \omega_i$ is described by 
\begin{eqnarray}
\hbar \omega_i= \sqrt {\frac{{{E_\text{ph}}}}{{2\varepsilon V}}} {\boldsymbol{\mu}_{i}}\cdot\boldsymbol{e } = -\hbar\omega_{0} \cos {\theta _i}, 
 \label{interaction energy}
\end{eqnarray}
where $\varepsilon$ and $V$ are the effective dielectric constant of the active layer and the effective mode volume of the cavity photon, respectively.
${\theta _i}$ is the angle between $\boldsymbol{\mu}$$_{i}$ and $\boldsymbol{e}$ (Fig.~\ref{model}(b)). 
$\omega_{0}$ equals $\sqrt {\frac{{{E_\text{ph}}}}{{2\varepsilon V}}} \frac{\mu}{\hbar}$. In this case, the state $\left|\mathscr{B}\right\rangle$ is not the eigenstate for the Hamiltonian, as confirmed by operating the Hamiltonian (Eqs.~(\ref{tc_H}) substituted by Eqs.~(\ref{interaction energy})) to the bright state (Eqs.~(\ref{bright states})).
 Instead, the state $\left|\mathscr{B}_{\rm{1D}}\right\rangle (\equiv \alpha \left| {g,1} \right\rangle + \beta \left| \rm{B_{1D},0} \right\rangle$) becomes an eigenstate (see SI B1).  
 $\left| \rm{B_{1D}}\right\rangle(\equiv {\left( {\sum\limits_{i = 1}^N {{{\cos }^2}{\theta _i}} } \right)^{ - \frac{1}{2}}}\sum\limits_{i = 1}^N {\cos {\theta _i}\left| {{e^i}} \right\rangle )}$ is a modification of the delocalized exciton state $\left| \rm{B}\right\rangle$, in which the probability density of each state $\left| {{e^i}} \right\rangle$ is proportional to $\cos ^{2} {\theta _i}$; we call the state $\left| \rm{B_{1D}}\right\rangle$ as an orientation-dependent delocalized state, because the probability density of the $i_{th}$ state depends on the orientation of the $i_{th}$ molecule. 
$\left| \rm{B_{1D},0} \right\rangle$ is the state where the molecules are in the state $\left| \rm{B_{1D}}\right\rangle$, and simultaneously, no cavity photon exists.
 The stationary Schr\"{o}dinger equation for the state $\left|\mathscr{B}_{\rm{1D}}\right\rangle$ can be simplified into the same equation as Eq.(\ref{matrix_2}), with the interaction energy $\hbar \Omega$ substituted by $\hbar \Omega_{\rm{1D}}$=$\sqrt {\sum\limits_{i = 1}^N {{{\cos }^2}{\theta _i}} } \hbar \omega _{0}$. The interaction energy equals $\sqrt {\frac{N}{3}} \hbar \omega_{0}$ when the orientation directions of the molecules are completely random under the condition that $N$ is large (see SI B2).
 The model described above is summarized in Fig.~\ref{model}(b).
 In the case that the $N$ molecules have a two-fold degenerate (2D) dipole composed of two orthogonal 1D dipoles and that their dipoles are the same strength, a similar discussion is possible by defining the orientation-dependent delocalized state for the 2D dipole $\left| \rm{B_{2D}} \right\rangle$ (see SI B3). The interaction energy becomes $\sqrt {\frac{2N}{3}} \hbar \omega_{0}$, when the orientation directions are random and $N$ is large. 
 
\subsubsection{2.2.3. Strong coupling between a photon and $N$ CdSe NPLs}
As discussed in \textbf{2.3.1}, the CdSe NPLs in the microcavity can be regarded approximately as two-level systems composed of the ground and hh-exciton states. In this subsection, we consider only the hh-exciton transition and ignore the lh-exciton transition.
The transition dipole of the hh-exciton in wurtzite CdSe NPLs is the 2D dipole of which the dipole strengths of the two orthogonal 1D dipoles are different. The orthogonal 1D dipoles are in the plane perpendicular to the [001] thickness direction of CdSe NPLs~\cite{Riccardo_2017} (see Fig. S\!\!~\ref{twodimensional}(a) in SI B3). The shapes of lateral directions of the NPLs are generally either rectangular or square~\cite{Bertrand_2016}. In the case of the rectangular NPLs, the dipole strength along the length direction is larger than that along the width direction, and the strength ratio depends on the length and width of the NPLs~\cite{Yoon_2017}.
The NPLs used in this study were rectangular; the aspect ratio was large, with a long length and a narrow width (Fig.~\ref{sample}(b)), which ensures that the dipole strength of the hh-exciton transition along the length direction is much stronger than that along the width direction. Thus, we can treat the transition dipole like a 1D dipole along the length direction, and we can expect strong coupling of the molecules having 1D dipoles in the microcavity, as long as disturbing effects on the strong coupling, such as cavity loss, exciton dumping, and NPL rotation, are small.
Note that the transition dipole of some of the representative colloidal NCs, such as CdSe QD~\cite{Chung_2003, Tani_2010} and CdSe NPLs in a square shape, can be regarded as a 2D dipole, of which the dipole strengths of the orthogonal 1D dipoles are the same. Thus, the model described in \textbf{2.2.2}, which treats the lowest excited state of the material system as the orientation-dependent delocalized states, is widely applicable for understanding the strong coupling and related optical properties of microcavities containing colloidal NC systems.
\subsubsection{2.2.4. Cavity photon energy in planar microcavities}
In planar microcavities, $E_{\rm{ph}}$ is given by
\begin{eqnarray}
{E_{\rm{ph}}} = \frac{{\hbar c}}{{{n}}}\left| {\vec k} \right| = \frac{{\hbar c}}{{{n}}}\sqrt {{{\left| {{{\vec k}_ \bot }} \right|}^2} + {{\left| {{{\vec k}_{/\!/}}} \right|}^2}},
 \label{wave vector}
\end{eqnarray}
where $c$ and $n$ are the speed of light and the refractive index in the microcavity.
${\vec k}$, ${{{\vec k}_ \bot }}$ and ${{{\vec k}_{/\!/}}}$ are the wave vectors of light in the microcavity, and the associated perpendicular and parallel components, respectively.
Especially in the microcavity designed for $\lambda$/2 mode, ${E_{\mathrm{ph}}}$ is written as
\begin{eqnarray}
{E_{\mathrm{ph}}}(\theta ) = \frac{{\hbar c}}{{{n}}}\sqrt {{{\left( {\frac{\pi }{L}} \right)}^2} + {{\left| {{{\vec k}_{/\!/}}} \right|}^2}} = \frac{\hbar c}{nL}{\left\{ {1 - {{\left( {\frac{{\sin \theta }}{{{n}}}} \right)}^2}} \right\}^{ - \frac{1}{2}}},
 \label{wave vector2}
\end{eqnarray}
where $L$ and $\theta$ are the cavity length and the external angle (viewing angle), respectively~\cite{Lidzey_2002, Oda_2009}.
\subsection{2.3. Optical properties of the microcavity containing hexane-dispersed CdSe NPLs}
\subsubsection{2.3.1. Reflection properties}
\begin{figure}[htb]
 \centering
 \includegraphics[width=6in]
      {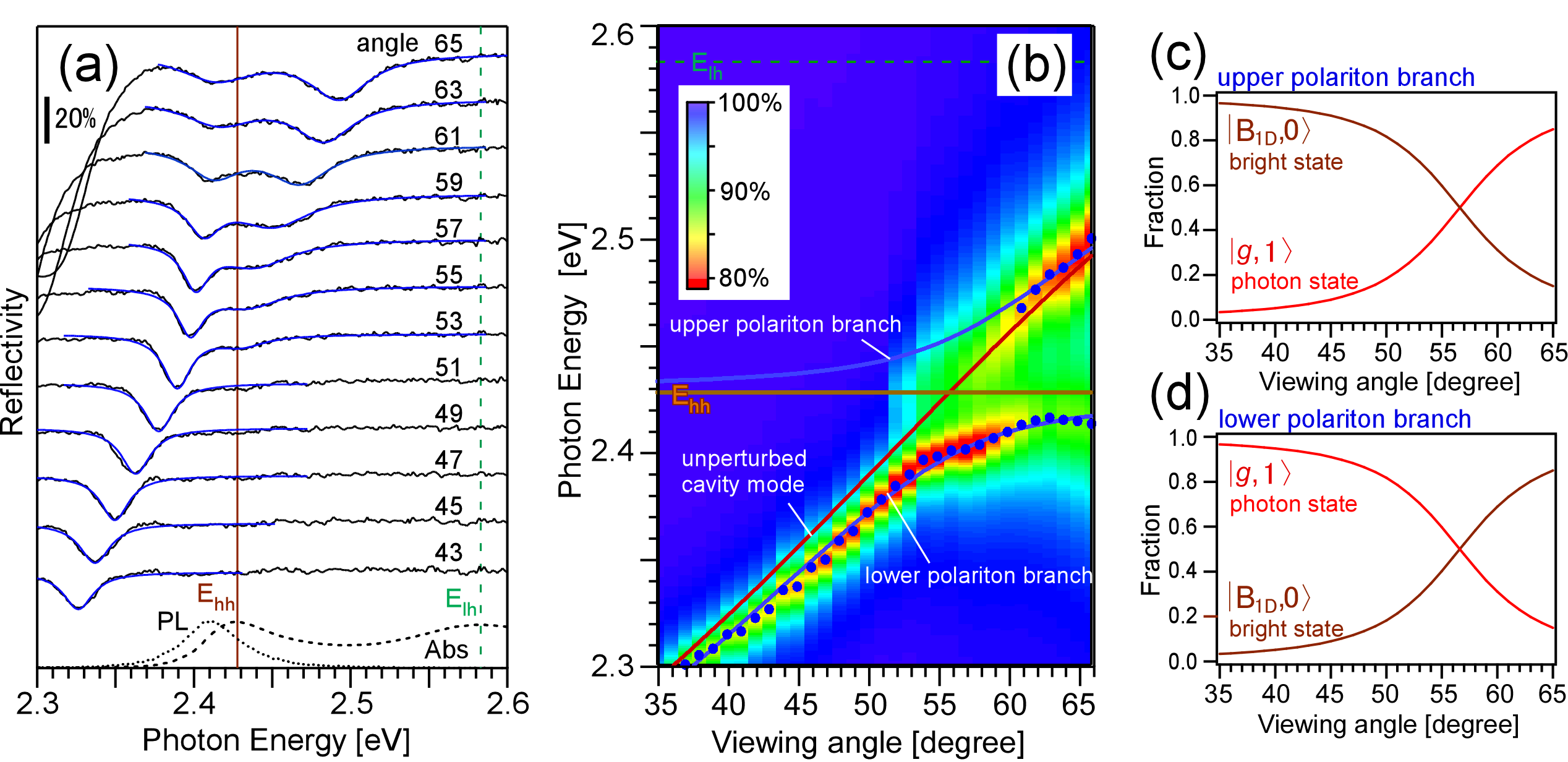}
 \caption{(a) Reflectivity spectra (black lines) and fitting curves (blue lines) of the microcavity for viewing angles of 43$^{\circ}$ to 65$^{\circ}$. (b) Dip energies of the fitting curves (solid circles) and dispersion curves obtained by least-squares fitting using Eq. (4)) (blue lines).   
 Mixing fraction of the photon state (red lines) and the orientation-dependent delocalized state (brown lines) for the upper (c) and lower (d) polariton branch, respectively. }
 \label{fig3}
\end{figure}

The black lines in Fig.~\ref{fig3}(a) show the transverse electric (TE) polarized reflectivity spectra of the microcavity as a function of the angle $\theta$$_{\rm{obs}}$ in the range 43$^\circ$-65$^\circ$.
Two dips appear in each spectrum in the range around 53$^\circ$-65$^\circ$, and they show anti-crossing behavior in the vicinity of $E_{\text{hh}}$ (the vertical brown line).
Only one dip is visible in the range around 43$^\circ$-51$^\circ$.
Note that an abrupt decrease, seen at the low energy side at higher angles (e.g., $\lesssim$ 2.37 eV at 65$^\circ$), corresponds to the low-energy edge of the cavity stopband of the microcavity. Each spectrum is fitted with the function of $R^{\theta_{\rm{obs}}} - \sum\limits_{i = 1}^N{L_{i}^{\theta_{\rm{obs}}}(E)}$ using a least squares method (LSM), where $R^{\theta_{\rm{obs}}}$ is a constant ($\simeq$100$\%$) and $L_{i}^{\theta_{\rm{obs}}}(E)$ is the $i^{th}$ Lorentzian curve at $\theta$$_{\rm{obs}}$ ($i$=2 for $\theta_{\rm{obs}}\geq$53$^\circ$, $i$=1 for $\theta_{\rm{obs}}<$53$^\circ$). The fitted curves reproduce the reflectivity spectra well, as shown by the blue lines in Fig.~\ref{fig3}(a).
Figure~\ref{fig3}(b) is an energy-angle ($E$-$\theta_{\rm{obs}}$) map of the reflectivity intensities, produced not from the experimental spectra but rather from the fitting curves to exclude the influence of the abrupt decrease at the low-energy side at high angles. The blue circles in Fig.~\ref{fig3}(b) indicate the dip energies estimated from the Lorentzian fittings. We fitted the dip energies with Eq.~(\ref{polariton energies}) using the LSM. The vacuum Rabi splitting energy 2$\hbar\Omega_{\rm{1D}}$, the refractive index $n$, and the cavity length $L$ were used as the fitting parameters, and the exciton energy $E_{\rm{ex}}$ was used as a fixed value (absorption peak energy of the black line in Fig.~\ref{sample}(a): $E_{\text{hh}}$=2.4278 eV).
The angle dependence of higher energy dips is fitted well by the curve of the upper polariton branch, and that of the lower energy dips is also fitted well by that of the lower polariton branch, as shown by the blue lines in Fig.~\ref{fig3}(b). 2$\hbar\Omega_{\rm{1D}}$ = 53.5 meV, $n$ = 1.91 and $L$ = 148 nm were evaluated.
In a study of microcavities containing J-aggregated cyanine dyes~\cite{J2010}, the evaluated refractive index in the microcavities was roughly the same as that of the matrix of the microcavities when the dye molecule density in the matrix was low, and increased as the density increased. Considering the increase, the evaluated $n$ and $L$ are thought to be reasonable for the $\lambda$/2 microcavity with a hexane solution layer of a refractive index of 1.4, containing the concentrated NPLs. Therefore, we conclude that the RT strong coupling with the vacuum Rabi splitting energy of 53.5 meV forms between a single photon and the orientation-dependent delocalized state of the NPLs in the $\lambda$/2 microcavity.
Note that the fitting for the upper polariton was applied to only the data at $\theta_{\rm{obs}}\geq$61$^\circ$ to increase the fitting accuracy, because the dips of the upper polariton at $\theta_{\rm{obs}}<$61$^\circ$ are small and broad, as shown in Fig.~\ref{fig3}(a) and (b). The origin of the broadness of the upper polariton near $E_{\mathrm{hh}}$ is not clear at present. It may be due to an asymmetry of the lower absorption band of the NPLs (Fig.~\ref{sample}(a)) or the influence of phonon coupling to the polariton states.~\cite{Herrera_2018}
Figure ~\ref{fig3}(c) and (d) display the mixing fraction of the bare photon (red line) and exciton (brown line) states, corresponding to $\left|\alpha\right|^2$ and $\left|\beta\right|^2$ in Eq.~(\ref{matrix_2}), in the upper and lower polariton branches, respectively. As seen in Fig.~\ref{fig3}(d), the lower branch is predominantly photonic in the low-angle region, changing its character from photonic to excitonic with increasing angle. Considering that incident light couples into the cavity via the photon component~\cite{prism_2011}, the reason that only a dip due to the lower polariton branch is visible at the low angle region and another dip due to the upper polariton branch becomes apparent with increasing angle (Fig.~\ref{fig3}(a)) can be understood as the dip area increases with increasing $\left|\alpha\right|^2$.
The Q-value of a microcavity is described by $\frac{{{E_{\rm{{cav}}}}}}{{\Delta {E_{\rm{cav}}}}}$, where $E_{{\rm{cav}}}$ and $\Delta{E_{{\rm{cav}}}}$ are the cavity photon energy and its broadness. We estimated the Q-value to be 116 by calculating the average $\frac{{E_{l}}}{\Delta E_{l}}$ in the low-angle range (33$^\circ$-38$^\circ$), where ${E_{l}}$ and $\Delta E_{l}$ are the dip energy of the lower polariton and its FWHM.
This is the approximation that the dip in the low-angle range is due to a pure photonic state; the red line of Fig.~\ref{fig3} in this angle range suggests that the approximation is reasonable.
To further investigate the factors governing the Q-value of our microcavity, we compared the Q-value estimated by a macroscopic PL measurement with the excitation spot $\phi$ $\simeq$ 0.6 mm (the same spot size and spatial resolution as that in the reflectance and PL measurements in this study) and that by a microscopic PL measurement with the excitation spot size $\phi$ $\simeq$ 30 $\mu$m at a low angle. The latter was much larger than the former; the Q-value of the latter measurement was 393 (see SI C1). Thus, we consider the Q-value estimated using the angle-dependent reflectance measurements in Fig.~\ref{fig3}(a) to be limited by small inhomogeneities of the cavity length $L$ in the spot size due to residual strain or tilt of the DBRs in forming the microcavity.
In this study, we use the light-matter coupling model based on the assumption that the NPLs are two-level systems. In contrast, another model that regards the NPLs as three-level systems, composed of a ground state, the hh-exciton state, and the lh-exciton state, has been used in the study of a microcavity containing colloidal NCs~\cite{AgNP_2016, prism_2011}. This model is practical for describing light-matter coupling in a microcavity of which the cavity mode’s energy crosses both the hh- and lh-exciton energies, although the treatment based on quantum mechanics is unclear in the case of nonuniformly orientated many-body NCs having the 1D (or 2D) transition dipole. We also applied the three-level model to our data to fit the dispersion relation for comparison (see SI B4) and calculated the mixing fractions. We confirmed that the difference between the three-level model and our model is negligible over a wide angle range. This is because the cavity mode’s energy does not reach the lh-exciton energy ($E_{\mathrm{lh}}$) in our microcavity. We consider that our model is better for describing the light-matter interaction in our microcavity because the model is a good approximation of the system, as it is based on quantum mechanical theory, and its intuitive understanding is possible as shown in Fig.~\ref{model}(b).
\subsubsection{2.3.2. Photoluminescence properties}
\begin{figure}[htb]
 \centering
 \includegraphics[width=6.5in]
      {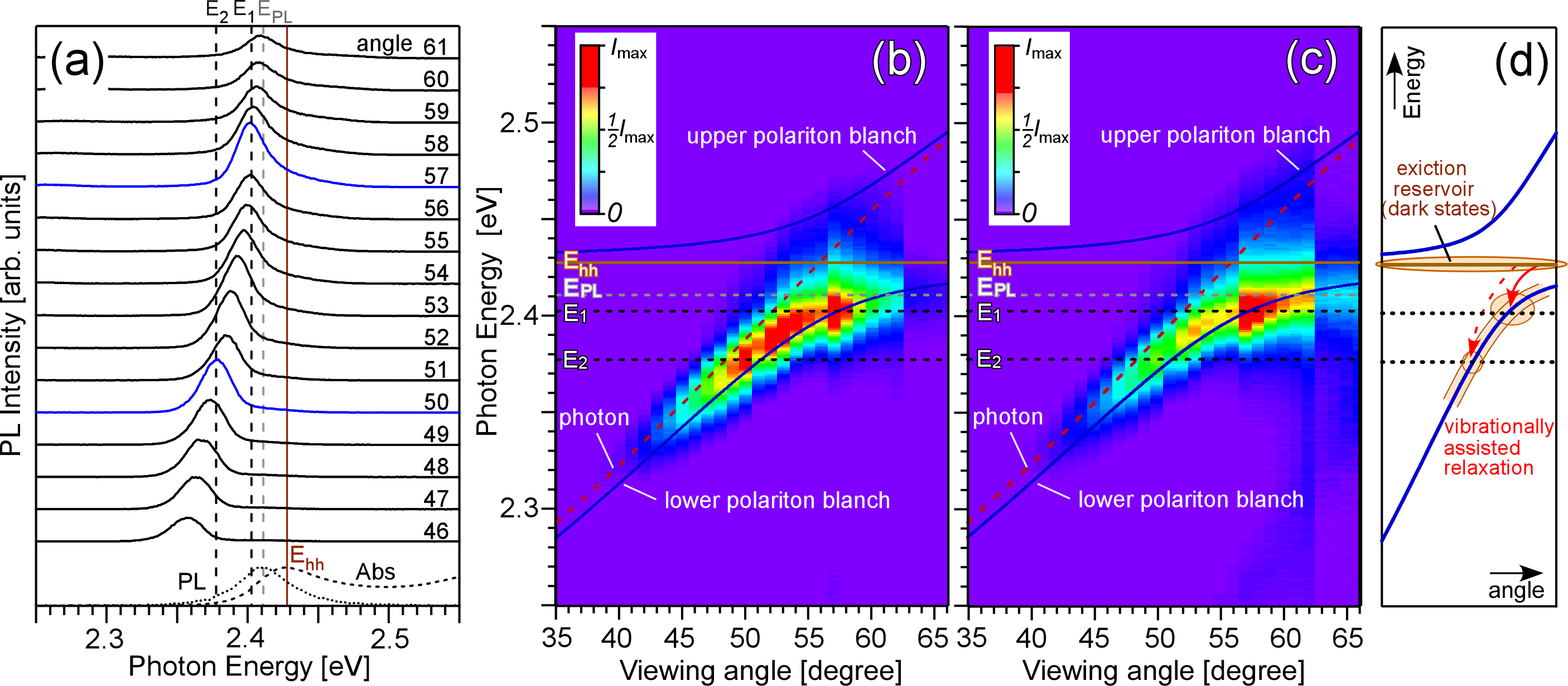}
 \caption{(a) PL spectra of the microcavity for viewing angles of 46$^{\circ}$ to 61$^{\circ}$ under 3.04-eV-light-irradiation (black and blue lines). Absorption peak energy due to the heavy-hole exciton state ($E_{\text{hh}}$) and PL peak energy ($E_{\text{PL}}$) of hexane-dispersed CdSe NPLs outside the microcavity are shown by vertical yellowish brown and gray broken lines, respectively. Vertical black broken lines show the energies $E_{n_{\mathrm{LO}}}$, i.e., the subtraction by the energy of $n_\mathrm{LO}$ LO phonon(s) ($n_\mathrm{LO}\hspace{1mm} \hbar \Omega_\mathrm{LO}$) from the energy of the dark states ($E_{\text{hh}}$). (b) Two-dimensional plot of PL spectra. Blue lines are the same dispersion curves in Fig. 3(b). (c) Two-dimensional plot of corrected PL spectra.}
 \label{fig4}
\end{figure}

The black and blue lines in Fig.~\ref{fig4}(a) show the TE polarized PL spectra measured at the angle of $\theta_{\rm{obs}}$ = 46$^\circ\sim $61$^\circ$ under non-resonant excitation (3.04 eV).
A single PL band is visible at each angle. Figure~\ref{fig4}(b) is an $E$-$\theta_{\rm{obs}}$ map of the PL intensities. The blue lines in Fig.~\ref{fig4}(b) are the same dispersion curves of the upper and lower polaritons shown by the blue lines in Fig.~\ref{fig3}(b). In the map, the high-intensity area of PL (red region) is distributed almost along the dispersion curve of the lower polariton, suggesting that the PL is due to the radiative recombination of the lower polariton. 
The PL signals from the upper polariton are almost invisible, which is similar to those of other microcavities containing colloidal NCs (NPLs~\cite{AgNP_2016, Qiu_2021, Yang_2022} and QDs~\cite{CdSeQD_2021}). The reason for the invisibility can be described based on the fast relaxation from the upper polariton to the dark states.
The PL spectra and map show the following features. (i) The high-intensity area of the lower polariton emission in the PL map is located in a relatively small angle range, whereas the low-intensity area in the reflectivity spectra, indicating the existence of the lower polariton state, is spread over a wide angle range, as shown in Fig.~\ref{fig3}(b). (ii) The PL energy in the high-intensity area is located slightly under the energy $E_\mathrm{PL}$ (broken gray lines in Fig.~\ref{fig4}(a) and (b)). (iii) The PL band areas at specific angles (57$^\circ$ and 50$^\circ$) are larger than that near the angles, as shown with the blue lines in Fig.~\ref{fig4}(a).
A similar phenomenon, corresponding to feature (iii), was reported in a study of a microcavity containing J-aggregated cyanine dyes dispersed in a dried PVA matrix~\cite{vib_assisted_David_2011}, although it has not been reported in NC microcavities. The PL band areas at some specific angles and neighboring angles in the J-aggregate microcavity were more significant than those around the angles. Coles et al.~\cite{vib_assisted_David_2011} revealed that the differences between the PL peak energies of the specific angles and the dark exciton energy ($E_{ex}$) in the microcavity well matched the vibrational energies $E_{\mathrm{vib}}^{\mathrm{J}}$ (= 40, 84, 150, and 184 meV) of the dye molecules estimated by Raman spectroscopy measurements. The investigation ultimately showed that the PL peaks were due to the emission of the lower polariton following the vibrational-assisted relaxation, accompanied by emitting phonons with energies of $E_{\mathrm{vib}}^{\mathrm{J}}$, from the dark states to the lower polariton state.
 
Meanwhile, it has been reported that one major and two minor phonon replica lines appear in the Raman spectrum of CdSe NPLs with a vibrational energy of $E_{n_{\mathrm{LO}}}$ (= $n_\mathrm{LO}\hspace{1mm} \hbar \Omega_\mathrm{LO}$ for $n_\mathrm{LO}$ = 1, 2, 3)~\cite{Raman}, where $\hbar \Omega_\mathrm{LO}$ (=25.1 meV) is the LO phonon energy of the CdSe NPLs.
We compared the vibrational energy $E_{n_{\mathrm{LO}}}$ and the difference between the dark state energy $E_{\mathrm{hh}}$ and the PL peak energies at specific angles. As a result, $E_{1}$ and $E_{2}$ well matched the differences at the specific angles, as shown by the black broken lines in Fig.~\ref{fig4}(a) and (b).
To further investigate the formation processes of the lower polariton, we estimated the polariton populations $D_\mathrm{\theta}$($E$) at each angle $\theta_ \mathrm{obs}$ using the relation $D_\mathrm{\theta}$($E$) = $\frac{P_\mathrm{\theta}(E)}{\alpha_\mathrm{\theta}^2}$ 
on the assumption that the PL intensity $P_\mathrm{\theta}$($E$)
is proportional to the product of the polariton population $D_\mathrm{\theta}$($E$) and the fraction of the photon component $\alpha_\mathrm{\theta}^2$ in the lower polariton state~\cite{AgNP_2016}. 
Figure~\ref{fig4}(c) shows the $E$-$\theta_{\mathrm{obs}}$ map of the polariton populations $D_\mathrm{\theta}$($E$). The photon fraction $\alpha_\mathrm{\theta}^2$ in the lower polariton state decreases as the angle $\theta_{\mathrm{obs}}$ increases, as shown in Fig.~\ref{fig3}(d). Thus, the high-intensity area of the polariton populations $D_\mathrm{\theta}$($E$) moves to a slightly higher angle side, and thus a slightly higher energy side, compared with that of the PL intensities $P_\mathrm{\theta}$($E$). However, the map of the polariton populations has the same features as that of the PL intensities (the intensity distribution of the $D_\mathrm{\theta}$($E$) near 50$^\circ$ is more visible in Fig. S6 of SI C2). Conversely, the high-intensity area of the polariton populations is more concentrated in the area with energy $E_{1}$, as shown in Fig.~\ref{fig4}(c). Thus, we conclude that the LO-phonon-assisted relaxation from the dark states to the lower polariton state is the primary process in forming the lower polariton, as illustrated by Fig. 4(c).
Based on the above results, we consider the relaxation process of the excited electrons in this microcavity as follows. When the excitation light with the energy of $E_\mathrm{exc}$ = 3.04 eV is illuminated on the microcavity at normal incidence, it excites the NPLs in the microcavity without reflection by the top and bottom DBR mirrors, because the $E_{\rm{exc}}$ is out of the energy range of the cavity stopband at the normal angle.
The excited state $\left| {{\tilde{e}}^{i},0} \right\rangle$ with energy $E_{\rm{exc}}$ is formed by light absorption, where the $\left| {{\tilde{e}}^{i},0} \right\rangle$ is the state that the $i^{th}$ molecule is in a high excited state (while the others are in their ground states) and, simultaneously, no cavity photon exists. Then, it relaxes immediately to the low-lying states, accompanied by phonon emission, similar to the relaxation of the high excited state within 2 ps in NPLs outside the microcavity~\cite{P.Sippel_2015}.
One of the low-lying states is the upper polariton. However, the PL was invisible, as shown in Fig.~\ref{fig4}(a) and (b), suggesting that the rate of subsequent relaxation to further low-lying states occurs much faster than that of the radiative recombination of the upper polariton.
This is consistent with the study on the J-aggregate microcavity~\cite{UP_2011} that the relaxation rate from the upper polariton to the dark states is ultrafast (150 fs).
Under the energy of the upper polariton state, there are ($N$-1)-fold degenerate dark states, as shown in Fig.~\ref{model} (b). The dark states are pure excitonic states, as discussed in \textbf{2.2.2}; thus, they are optically forbidden.
Besides, the relaxation rate from the upper to the dark states is much faster than that from the dark to the lower polariton states when the relaxation occurs owing to some perturbation. Thus, the dark states function as an exciton reservoir.
The difference in the relaxation rate is due to that of the density of the final state of each relaxation transition, because the transition rate due to a perturbation is proportional to the density of the final state according to Fermi’s golden rule~\cite{polariton_chemistry, Dirac_1927}; the density of the dark states is much denser than the lower polariton state due to the high degeneration of the dark states, which results in a significant difference in the relaxation time.
We showed that the high-intensity area of the lower polariton populations is concentrated at the energy of $E_{1}$ (Fig.~\ref{fig4}(c)). This is direct evidence that a pathway exists from the exciton reservoir to the lower polariton state due to a perturbation and that the perturbation causing the relaxation is the interaction between the exciton in the dark states and the LO-phonon. Figure~\ref{fig4}(c) also indicates that the LO-phonon-assisted relaxation is the dominant process for forming the population distribution of lower polaritons and that the contribution of the other relaxation processes, e.g., intra-relaxation in the lower polariton branch and direct relaxation from the high excited state $\left| {{\tilde{e}}^{i},0} \right\rangle$ to the lower polariton branch, are less critical than that of the LO-phonon-assisted relaxation.
There seem to be two more pathways to exit from the exciton reservoir in addition to the LO-phonon-assisted relaxation: one is a non-radiative annihilation to the ground state $\left| {g,0} \right\rangle$, where all molecules are in their ground states and no cavity photon exists, and the other is the pathway from the dark to the upper polariton states due to phonon absorption. However, we consider that both pathways have less influence on the relaxation dynamics. The time constant of non-radiative annihilation is thought to be slow ($>$ 1 ns) because the NPLs in solutions at RT usually show a high quantum yield of PL with a slow radiative time ($\left\langle \tau \right\rangle$=16 ns~\cite{Yadav_2017}). The time constant of the LO-phonon-assisted relaxation of the NPLs is not presently apparent, but the typical time of phonon-related relaxation of molecule systems is $\leq$10 ps~\cite{Herrera_2018}. Thus, it is thought that most of the excitons in dark states relax by LO-phonon-assisted relaxation before relaxing via non-radiative annihilation. Meanwhile, we can ignore the latter pathway to the upper state because it returns back to the dark states immediately. 
After the lower polariton is formed mainly due to the LO-phonon-assisted relaxation, it further relaxes to the ground state $\left| {g,0} \right\rangle$ by emitting a photon via the photon component with the energy of the lower polariton. The emitted photon is the PL of the microcavity. The lower polariton also has the relaxation pathway due to non-radiative annihilation, but it is less critical in this microcavity due to its slow time constant. 
\section{3. Summary and outlook}
We have developed a method for fabricating the $\lambda$/2 microcavity with a pair of DBR mirrors containing liquid-dispersed CdSe NPLs and characterized the optical properties of the microcavity based on the angle-dependent reflectivity and PL measurements operating at RT. Based on the reflectance measurements, strong light-matter coupling is achieved in the microcavity, and the Rabi-splitting energy is 53.5 meV. From PL measurements, the PL of the lower polariton is followed by the LO-phonon-assisted relaxation from the dark states to the lower state. We describe the reflectance and PL properties based on the model considering the coupling between the cavity photon and the delocalized exciton state over randomly oriented NPLs. 
In this paper, consider a microcavity in which the active layer comprises a solution. The fabrication technique and knowledge of the optical properties can extend to various colloidal NCs. Furthermore, mixing different types of NCs and using conjugates between NCs and other NCs (or NCs/dyes) are easily possible. Thus, we believe this study will contribute not only to new developments in lasing technology but also future investigations in polariton physics and chemistry.
\begin{acknowledgement}
The authors thank N. Wakayama of the Center for Instrumental Analysis of Kyushu Institute of Technology for his technical assistance in transmission electron microscopy analysis. We are grateful to T. Hosokawa, H. Matsuo, K. Nanami, K. Nagaki, K. Nakaishi, and K. Fujii for their collaboration in the early stage of this work. We acknowledge the financial support of Japan Society for the Promotion of Science KAKENHI Grant Nos. 16K04878 and 19K05197.
\end{acknowledgement}
\begin{suppinfo}
\renewcommand{\figurename}{Figure S\!\!}
\setcounter{figure}{0}
\subsection{A. Synthesis and characterization}
\subsubsection{A1. Synthesis and purification of 4-ML NPLs}
We synthesized 4-monolayer (ML) CdSe nanoplatelets (NPLs) based on the method of Bertrand et al.~\cite{Bertrand_2016}. First, we prepared cadmium myristate, as a precursor of the NPLs. Cadmium nitrate tetrahydrate in the amount of 0.6 g (99.0\%, Sigma-Aldrich, St. Louis, MO, USA) was dissolved in 25 mL of methanol (99.8\%, Wako Chemicals, Osaka, Japan). Separately, 1 g of natrium myristate (98.0\%, Wako Chemicals) was dissolved in 50 mL of methanol. By mixing the two solutions, the cadmium myristate was precipitated. We washed the precipitate with methanol three times and dried it in a vacuum for 1 day.
Then, 150 mg of cadmium myristate and 24 mg of selenide powder (99.99\%, Sigma-Aldrich) were mixed with 15 mL of octadecene (95.0\%, Sigma-Aldrich) in a 100-mL three-necked flask. After the mixture was degassed under vacuum at 110$^\circ$C for 20 min, it was placed under an argon atmosphere and stirred at 140$^\circ$C for 20 min. It was heated at the rate of 25$^\circ$C per minute, and 80 mg of cadmium acetate dihydrate (99.9\%, Wako Chemicals) was added at 210$^\circ$C. After the temperature reached 240$^\circ$C, it was held at this temperature for 20 min to grow the NPLs. Then, it was cooled to 160$^\circ$C, and 2 mL of oleic acid (99\%, Sigma-Aldrich) was added.
Finally, we purified the CdSe NPLs. The solution was left at room temperature for over 2 h. After discarding the supernatant, 15 mL of hexane (96.0\%, Wako Chemicals) solution was added to disperse the NPLs. The solution was centrifuged at 14,800 rpm for 30 min, and the supernatant was discarded. Hexane (15 mL) was added to the precipitation and centrifugated at 14,800 rpm for 30 min. The supernatant contained the 4-ML CdSe NPLs.
\subsubsection{A2. Shape and size of the NPLs}
Figure~S\ref{TEML} shows a transmission electron microscopy wide-field image of the NPLs. The NPLs consisted of two components. The main component had a rectangular shape with an average length and width of 13.5 $\pm$ 0.5 nm and 2.5 $\pm$ 0.1 nm, respectively. The other component had a roughly square shape with sides of 10-15 nm in length/width. The mixing ratio of the components was 77\% to 23\%.
\begin{figure}[htb]
 \centering
 \includegraphics[width=3.25in]
      {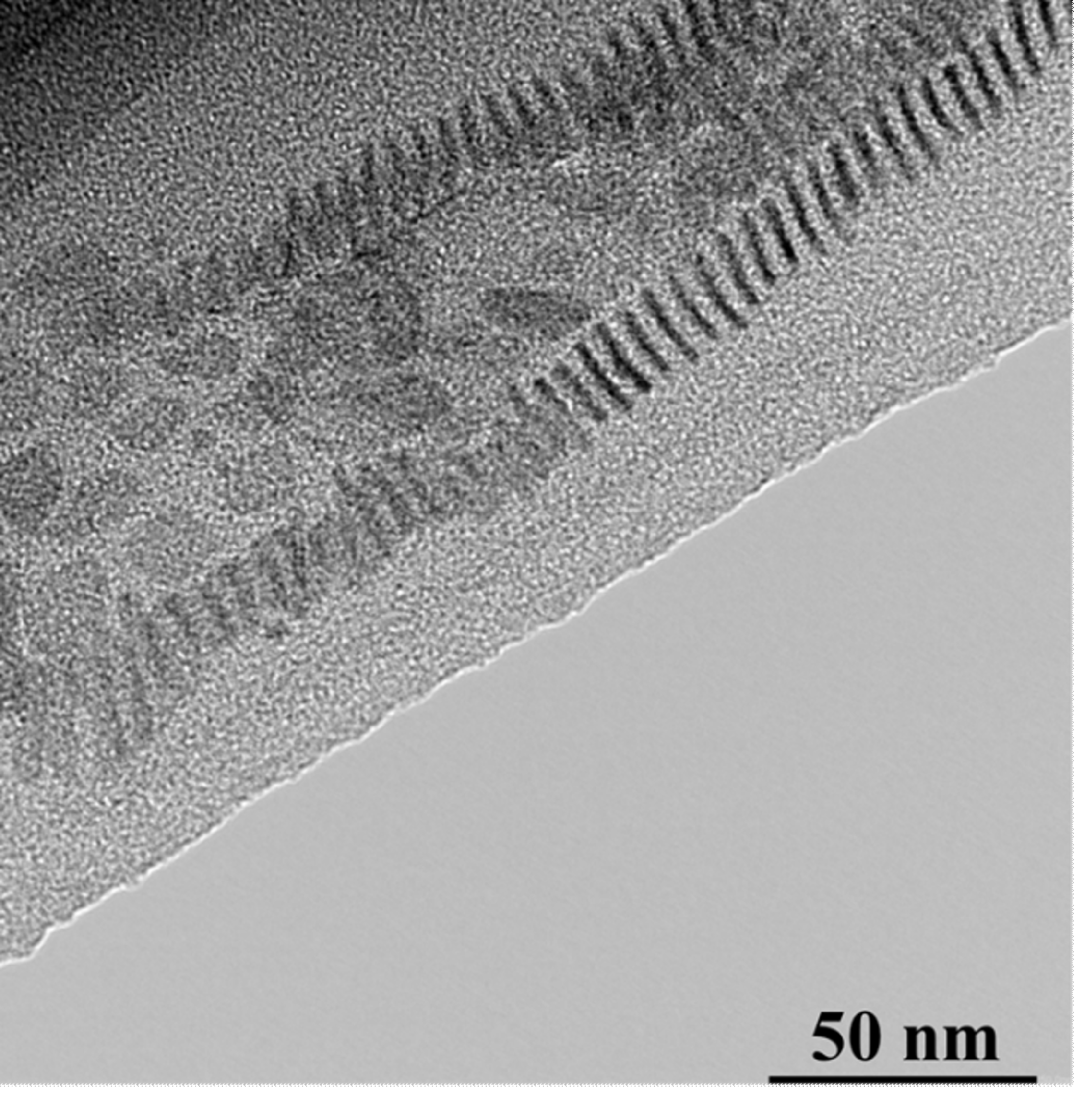}
  \caption{Transmission electron microscopy image of CdSe nanoplatelets (NPLs) taken with a wide field}.
 \label{TEML}
\end{figure}
\subsubsection{A3. Concentrated NPL solution for the liquid-dispersed microcavity}
We repeated the synthesis and purification processes described in A1 three times, and the sum solution was concentrated as described in the main text. The insets (a) and (b) in Fig.~S\ref{concentrated} show the pictures before and after concentration, respectively. 
To check damage caused by the concentration step, we prepared a re-diluted solution of the concentrated solution. Figure S2 shows the absorption and PL spectra of the NPLs in hexane before (black lines) and after (red lines) concentration. Note that the latter is the re-diluted solution and that all the spectra were normalized. We confirmed that the concentration process caused no spectral changes.
The sample used in this measurement is not the same as that used in the study in the text, but the method of synthesis and concentration are the same.
\begin{figure}[htb]
 \centering
 \includegraphics[width=4 in]
      {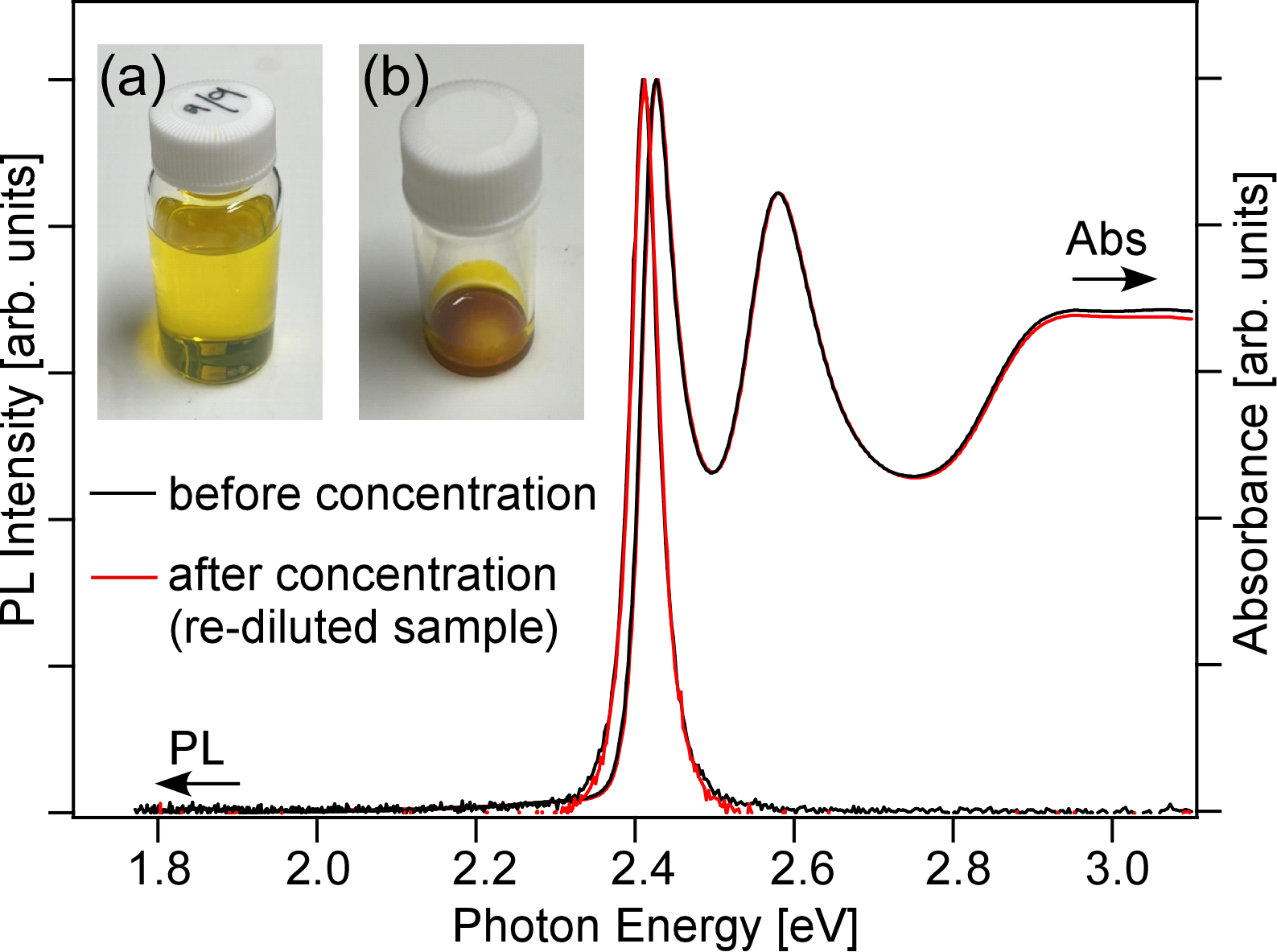}
  \caption{Insets show images of the NPLs (a) before and (b) after the concentration process. The black lines show the absorption and PL spectra of the NPLs in hexane before concentration. The red lines show those after concentration (re-diluted NPL solution). }.
 \label{concentrated}
\end{figure}
\subsection{B. Strong light-matter coupling}
\subsubsection{B1. Strong coupling between a photon and an orientation-dependent delocalized state $\left| {B_{\rm{1D}},0} \right\rangle$ of N molecules having a 1D dipole}
We consider the strong coupling between a linearly polarized photon and $N$ molecules having 1D transition dipole moments $\boldsymbol{\mu}_{i}$ with the same magnitude $\mu$ but anisotropic in direction. Hamiltonian $H\rm{_{1D}}$ of the system in the rotating-wave approximation is given by
\begin{equation}
H \rm{_{1D}} = \it{E}_{\rm{ph}}{a^\dag }a + \sum\limits_{i = 1}^N {{\it{E}_{\rm{ex}}}\sigma _ + ^i\sigma _ - ^i} + \sum\limits_{i = 1}^N {\hbar \omega_\mathrm{{0}} \cos {\theta _i}\left( {{a^\dag }\sigma _ - ^i + a\sigma _ + ^i} \right)}. \tag{S1}
 \label{Hamiltonian_2}
\end{equation}
Here, we assume that the wavefunctions $\left| \Phi \right\rangle$=$ \alpha \left| {g,1} \right\rangle + \beta \left| {B_{\rm{1D}},0} \right\rangle $ are the eigenstates of the Hamiltonian $H\rm{_{1D}}$. By considering that the creation and annihilation operators satisfy the relations $
{a^\dag }\left| 0 \right\rangle = \left| 1 \right\rangle, a\left| 1 \right\rangle = \left| 0 \right\rangle, \sigma _ + ^i\left| {g} \right\rangle = \left| {e_i} \right\rangle, \sigma _ - ^i\left| {e_i} \right\rangle = \left| {g} \right\rangle$, then $H_{\mathrm{1D}}\left| \Phi \right\rangle$ can be expressed by the sum of the following three terms:
\begin{equation}
{E_{\rm{ph}}}{a^\dag }a\left| \Phi \right\rangle = \cdots = \alpha {E_{\rm{ph}}}\left| {g,1} \right\rangle, \tag{First term}
 \label{First term}
\end{equation}
\begin{equation}
\sum\limits_{i = 1}^N {{E_{\rm{ex}}}\sigma _ + ^i\sigma _ - ^i} \left| \Phi \right\rangle = \cdots = \beta {E_{\rm{ex}}}\left| {B_{\rm{1D}},0} \right\rangle, \tag{Second term}
 \label{Second term}
\end{equation}
\begin{equation}
\sum\limits_{i = 1}^N {\hbar \omega_{0} \cos {\theta _i}\left( {{a^\dag }\sigma _ - ^i + a\sigma _ + ^i} \right)} \left| \Phi \right\rangle = \cdots = \hbar \Omega_{\mathrm{1D}} \left( {\beta \left| {g,1} \right\rangle + \alpha \left| {B_{\rm{1D}},0} \right\rangle } \right), \tag{Third term}
 \label{Third term}
\end{equation}
where the $\hbar \Omega_{\mathrm{1D}} \equiv \hbar \omega_{0} \sqrt {\sum\limits_{i = 1}^N {{{\cos }^2}{\theta _i}} }$ is the interaction energy of this system. Thus, the Schr\"{o}dinger equation $H_{\mathrm{1D}}\left| \Phi \right\rangle$ = $E^{\mathrm{1D}}_{u,l}$ $ \left| \Phi \right\rangle$ is rewritten as
\begin{equation} \left( {\begin{array}{*{20}{c}}
{{E_{ph}}}&{\hbar \Omega_{\mathrm{1D}}}\\
{\hbar \Omega_{\mathrm{1D}}}&{{E_{ex}}}
\end{array}} \right)\left( {\begin{array}{*{20}{c}}
\alpha \\
\beta 
\end{array}} \right) = E^{\mathrm{1D}}_{u,l}\left( {\begin{array}{*{20}{c}}
\alpha \\
\beta 
\end{array}} \right).\tag{S2}
 \label{equation}
\end{equation}
Equation (S2) is the same as Eq.(3) except for the interaction energy $\hbar \Omega_{\mathrm{1D}}$. This means that the wavefunctions $\left| \Phi \right\rangle$=$ \alpha \left| {g,1} \right\rangle + \beta \left| {B_{\rm{1D}},0} \right\rangle $ $(\equiv\left|\mathscr{B}_{\mathrm{1D}} \right\rangle)$ are the bright states and the energies $E^{\mathrm{1D}}_{u,l}$ are the upper and lower polariton energies in the N-molecule system having a 1D dipole.
\subsubsection{B2. Interaction energy between a photon and the state $\left| {B_{\mathrm{1D}},0} \right\rangle$ of randomly orientated N molecules having 1D dipole}
The probability density that the vector starting at the origin and passing through a surface element $d{\theta}\sin {\theta} d{\phi}$ on a unit sphere centered at the origin is described as
\begin{equation}
\frac{\rm{(Area{\rm{ }}\:of\:the\:{\rm{ surface}}\:element)}}{\rm{(Surface{\rm{ }}\:area\:of\:the\:unit\:{\rm{ sphere)}}}} =\frac{d{\theta }\sin {\theta } d{\phi }} {4\pi } \tag{S3}
 \label{ probability density }
\end{equation}
Thus, the expectation value of $\overline {{{\cos }^2}{\theta _i}}$ under the condition that the orientation direction of the N molecules is
random and that N is large is given by
\begin{equation}
\overline {{{\cos }^2}{\theta _i}} = \int_0^\pi {\int_0^{2\pi } {{{\cos }^2}{\theta _i}} } \cdot \frac{{d{\theta _i}\sin {\theta _i}d{\phi _i}}}{{4\pi }} = \frac{1}{{4\pi }}\int_0^\pi {{{\cos }^2}{\theta _i}\sin \theta _i d{\theta _i}\int_0^{2\pi } {d{\phi _i}} } = \frac{1}{3}
\tag{S4}
 \label{ave_cos}
\end{equation}
Therefore,
\begin{equation}
\hbar \Omega_{\mathrm{1D}}^{\mathrm{random}} = \sqrt {\sum\limits_{i = 1}^N {\overline {{{\cos }^2}{\theta _i}} } } \hbar {\omega _0} = \sqrt {\frac{N}{3}} \hbar {\omega _0}\
\tag{S5}
 \label{1D}
\end{equation}
 \subsubsection{B3. Strong coupling and the interaction energy between a photon and the $\left| {B_{\rm{2D}},0} \right\rangle$ of N molecules having a 2D dipole}
We consider the strong coupling between a photon polarizing with the oscillation direction along the unit vector $\textit{\textbf{e}}$ (z-axis direction in Fig.~S\ref{twodimensional}(b) and (c)) and the $N$ molecules having a 2D transition dipole $\boldsymbol{\mu}$$_{i}$ (Fig.~S\ref{twodimensional}(a)) with the same magnitude ${\mu _i} = \mu$ but with anisotropic directions. 
The strength of the light-matter interaction of the $i^{th}$ molecule $\hbar \omega_i$ is described by $ \sqrt {\frac{{{E_\text{ph}}}}{{2\varepsilon V}}} {\boldsymbol{\mu}_{i}}\cdot\boldsymbol{e} = -\sqrt {\frac{{{E_\text{ph}}}}{{2\varepsilon V}}} \mu \cos(\frac{\pi}{2}-{\xi _i} ) = -\hbar\omega_{0} \sin {\xi _i}$, where ${\xi _i}$ is the angle between $\boldsymbol{e}$ and the normal vector from the origin of the 2D dipole plane of the $i^{th}$ molecules (Fig.~S\ref{twodimensional}(b)). The Hamiltonian $H\rm{_{2D}}$ is given by
\begin{equation}
H \rm{_{2D}} = \it{E}_{\rm{ph}}{a^\dag }a + \sum\limits_{i = 1}^N {{\it{E}_{\rm{ex}}}\sigma _ + ^i\sigma _ - ^i} + \sum\limits_{i = 1}^N {\hbar \omega_\mathrm{{0}} \sin {\xi _i}\left( {{a^\dag }\sigma _ - ^i + a\sigma _ + ^i} \right)}. \tag{S6}
 \label{Hamiltonian_3}
\end{equation}
The difference between $H \rm{_{2D}}$ and $H \rm{_{1D}}$ is only the replacement of cos ${\theta _i}$ with sin ${\xi _i}$.
We assume that the state $\left| \Phi \right\rangle$ $=\alpha \left| {g,1} \right\rangle + \beta \left| \rm{B_{2D},0} \right\rangle$ ($\equiv \left|\mathscr{B}_{\rm{2D}}\right\rangle$) satisfies Eq.(\ref{matrix_2}). We define the $\left| \rm{B_{2D},0} \right\rangle$ as the state where the molecules are in the orientation-dependent delocalized state for the 2D dipole $\left| \rm{B_{2D}}\right\rangle$ $(\equiv {\left( {\sum\limits_{i = 1}^N {{{\sin }^2}{\xi _i}} } \right)^{ - \frac{1}{2}}}\sum\limits_{i = 1}^N {\sin {\xi _i}\left| {{e^i}} \right\rangle })$ and no cavity photon exists. Then, the difference between $\left| \rm{B_{2D},0} \right\rangle$ and $\left| \rm{B_{1D},0} \right\rangle$ is also only the replacement of cos ${\theta _i}$ with sin ${\xi _i}$. Thus, similar calculations and arguments in B1 hold in this case.
This means that the states $\left|\mathscr{B}_{\rm{2D}}\right\rangle$ are the bright states for the Hamiltonian $H \rm{_{2D}}$. The interaction energy $\hbar \Omega_\mathrm{{2D}}$ is $\sqrt {\sum\limits_{i = 1}^N {{{\sin }^2}{\xi _i}} } \hbar \omega_{0}$ in the 2D case.
The expectation value of $\overline {{{\sin }^2}{\xi _i}}$ under the condition that the orientation direction of the N molecules is random and that N is large is calculated as follows:
\begin{equation}
\overline {{{\sin }^2}{\xi _i}} = \int_0^\pi {\int_0^{2\pi } {{{\sin }^2}{\xi _i}} } \cdot \frac{{d{\xi _i}\sin {\xi _i}d{\eta _i}}}{{4\pi }} = \frac{1}{{4\pi }}\int_0^\pi {{{\sin }^3}{\xi _i} d{\xi _i}\int_0^{2\pi } {d{\eta _i}} } = \frac{2}{3}.
\tag{S7}
 \label{ave_sin}
\end{equation}
Therefore,
\begin{equation}
\hbar \Omega_{\mathrm{2D}}^{random} = \sqrt {\sum\limits_{i = 1}^N {\overline {{{\sin }^2}{\xi _i}} } } \hbar {\omega _0} = \sqrt {\frac{2N}{3}} \hbar {\omega _0}.\
\tag{S8}
 \label{2D}
\end{equation}
 
 \begin{figure}[htb]
 \centering
 \includegraphics[width=4 in]
      {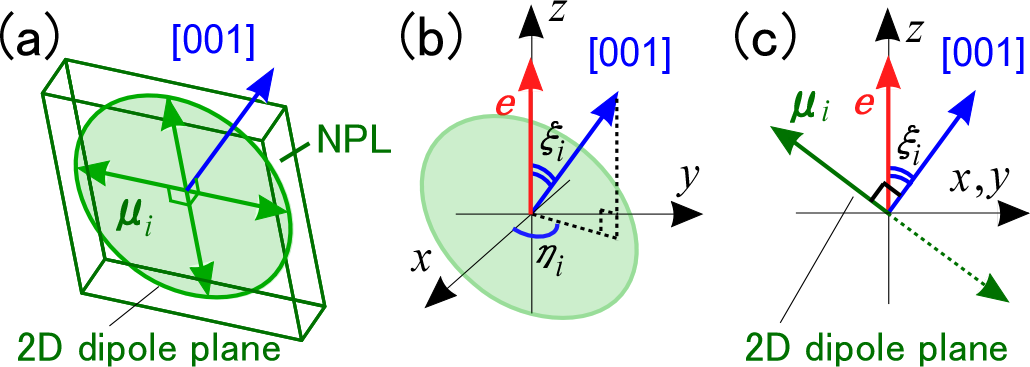}
 \caption{(a) 2D dipole plane of the wurtzite CdSe NPLs. (b) and (c) show the relationship between the 2D dipole plane and the oscillation direction of light.${\xi _i}$ and ${\eta _i}$ are the polar and azimuthal angles, respectively.} 
\label{twodimensional}
\end{figure}
\subsubsection{B4. Strong coupling between a photon and a molecule with a three-level system}
Here, we consider the light-matter strong coupling in microcavities between a photon in a single mode and a molecule of the three-level system, i.e., a ground state, a ${E_{{\rm{hh}}}}$ state, and a ${E_{{\rm{lh}}}}$ state. This model is one of models used most often for a microcavity containing colloidal NCs to analyze the data~\cite{AgNP_2016, prism_2011}. The stationary Schr\"{o}dinger equation is given by
\begin{equation}
\left( {\begin{array}{*{20}{c}}
{{E_{{\rm{ph}}}}}&\hbar\gamma &\hbar\delta \\
\hbar\gamma &{{E_{{\rm{hh}}}}}&0\\
\hbar\delta &0&{{E_{{\rm{lh}}}}}
\end{array}} \right)\left( {\begin{array}{*{20}{c}}
\alpha \\
\beta \\
\chi 
\end{array}} \right) = {E_{\rm{P1,\,P2,\,P3}}}\left( {\begin{array}{*{20}{c}}
\alpha \\
\beta \\
\chi 
\end{array}} \right),  \tag{S9}
 \label{matrix_3}
\end{equation}
where $\hbar\gamma$ ($\hbar\delta$) is the vacuum Rabi splitting energy between a photon and an hh-exciton (lh-exciton). $E_{\rm{P1,\,P2,\,P3}}$ are the energies of the light-matter hybrid states corresponding to a lower, a middle, and an upper polariton, respectively. The light-matter hybrid states are described by linear couplings of the photon, hh-exciton, and lh-exciton states, respectively. $\alpha$, $\beta$, and $\chi$ are the coefficients of the linear coupling.
We fitted the dip energies shown by the blue circles in Fig.~\ref{fig3}(b) with the energies of $E_{\rm{P1,\,P2}}$ using the least squares method on the assumption that the vacuum Rabi splitting energy $\hbar\gamma$, the refractive index $n$, and the cavity length $L$ are fitting parameters and that the exciton energies $E_{\rm{hh}}$ and $E_{\rm{lh}}$ are fixed values (absorption peak energies of the black line in Fig.~\ref{sample}(a): $E_{\text{hh}}$ = 2.4278 eV and $E_{\text{lh}}$ = 2.5839 eV). 
In the case of our microcavity, the cavity photon energy (the red line in Fig.~\ref{fig3}(b)) did not reach $E_{\rm{lh}}$, even at high viewing angles; thus, there is less information to determine the interaction energy $\hbar\delta$. As such, we use the relation $\hbar\delta$ = 0.94$\hbar\gamma$ in the literature~\cite{AgNP_2016} to reduce the ambiguity.
The broken black lines, denoted by P1, P2, and P3 in Fig. S\!\!~\ref{fig4}(a), indicate the fitted curves corresponding to $E_{\rm{P1,\,P2,\,P3}}$. The solid blue lines indicate the curves of the upper and lower polariton branches shown in Fig.~\ref{fig3}(b). The curves of the P1 and the lower polariton branches are the same over the all-angle range. A small difference between that of P2 and the upper polariton branches is visible only at the higher angles ($\theta_{\rm{obs}}>$70$^\circ$).
The broken lines in Fig. S\!\!~\ref{fig4}(b) and (c) show the mixing fraction of the bare photon, hh-, and lh-exciton states for P2 and P1, respectively. The mixing fractions of the P1 branch look almost the same as those of the lower polariton branch. The mixing fractions of the P2 branch look almost the same as those of the upper polariton branch up to approximately 65$^\circ$. From these results, we consider that the contribution of the lh-exciton state to the optical properties is small in our microcavity.
It is noted that adopting the three-level system to our model that describes the interaction between a photon and nonuniformly oriented $N$ molecules having a 1D (or 2D) transition dipole moment is not simple. Thus, this extension is a future consideration.
 \begin{figure}[htb]
 \centering
 \includegraphics[width=4.5 in]
      {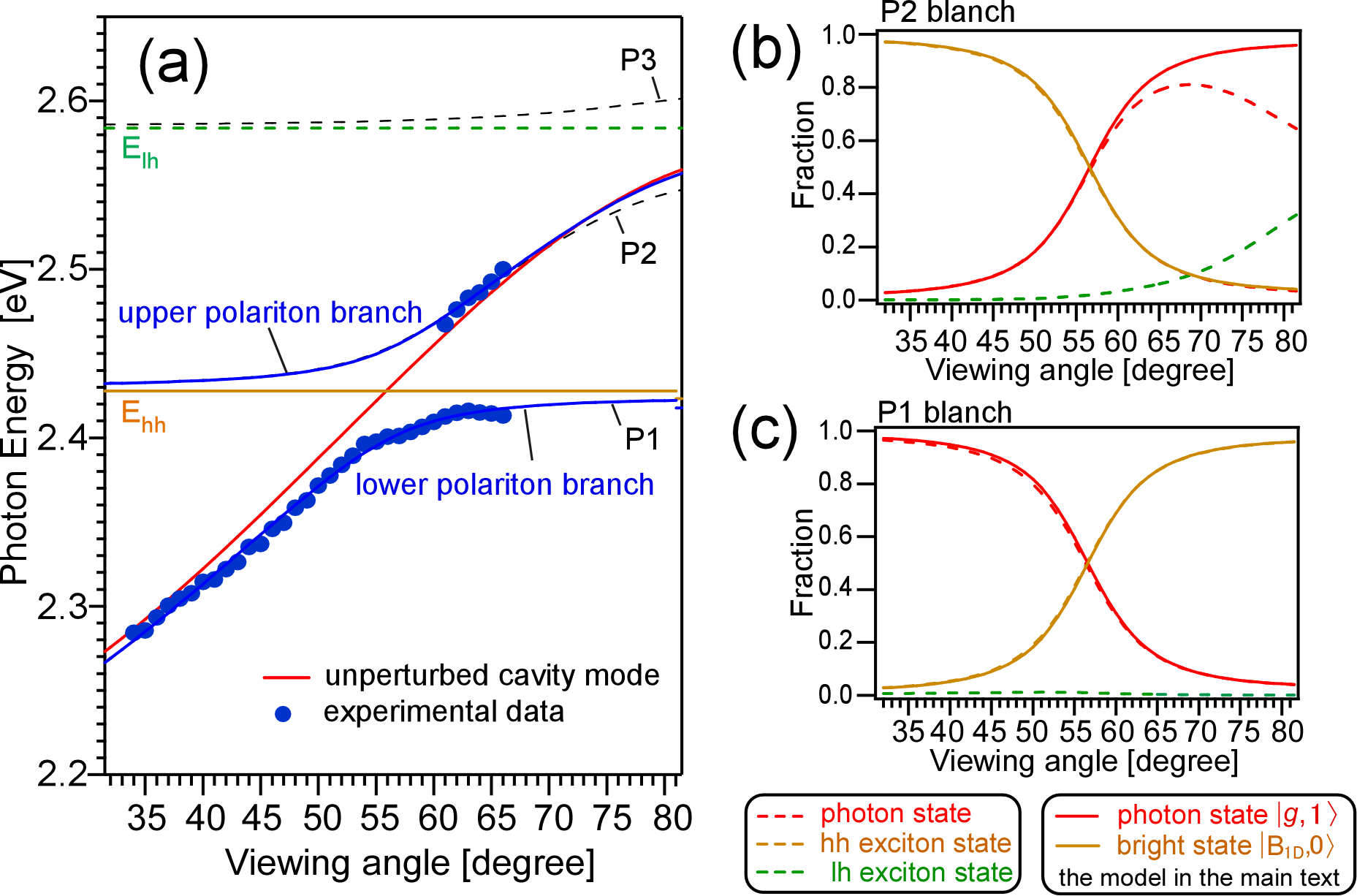}
  \caption{(a) Dip energies shown in Fig.~\ref{fig3}(a) (solid circles) and dispersion curves of the polariton states obtained by the least-squares fitting with the energies of $E_{\rm{P1,\,P2,\,P3}}$ (broken black lines). Solid blue lines are the dispersion curves of the upper and lower polariton branches with energies of $E_{u, l}$, as shown in Fig.~\ref{fig3}(b). Mixing fraction of photon (red broken lines), hh-exciton (brown broken lines), and light-hole exciton (green broken lines) states for (b) P2 and (c) P1 branches, respectively. Solid lines show the mixing fraction for the upper and lower polariton branches in Fig~\ref{fig3}(c) and (d).  
} 
  \label{Sreflectance}
\end{figure}
\subsection{C. Optical properties of the microcavity}
\subsubsection{C1. Quality factor estimated by a microscopic PL observation}
The red and blue lines in Fig.~S\ref{C1} show the PL spectra at $\theta_{\rm{obs}}$ = 35$^\circ$, of which the excitation spot sizes are $\phi$ $\simeq$ 0.6 mm (the same spot size as the reflectance and PL measurements shown in the main text) and $\phi$ $\simeq$ 30 $\mu$m, respectively. The full width at half maximum of the latter spectrum is five times narrower than that of the former. The quality factor Q-values were estimated as 79 and 393, respectively. Note that the sample used in this experiment differs from that used in the main text.
 \begin{figure}[htb]
 \centering
 \includegraphics[width=3.2 in]
      {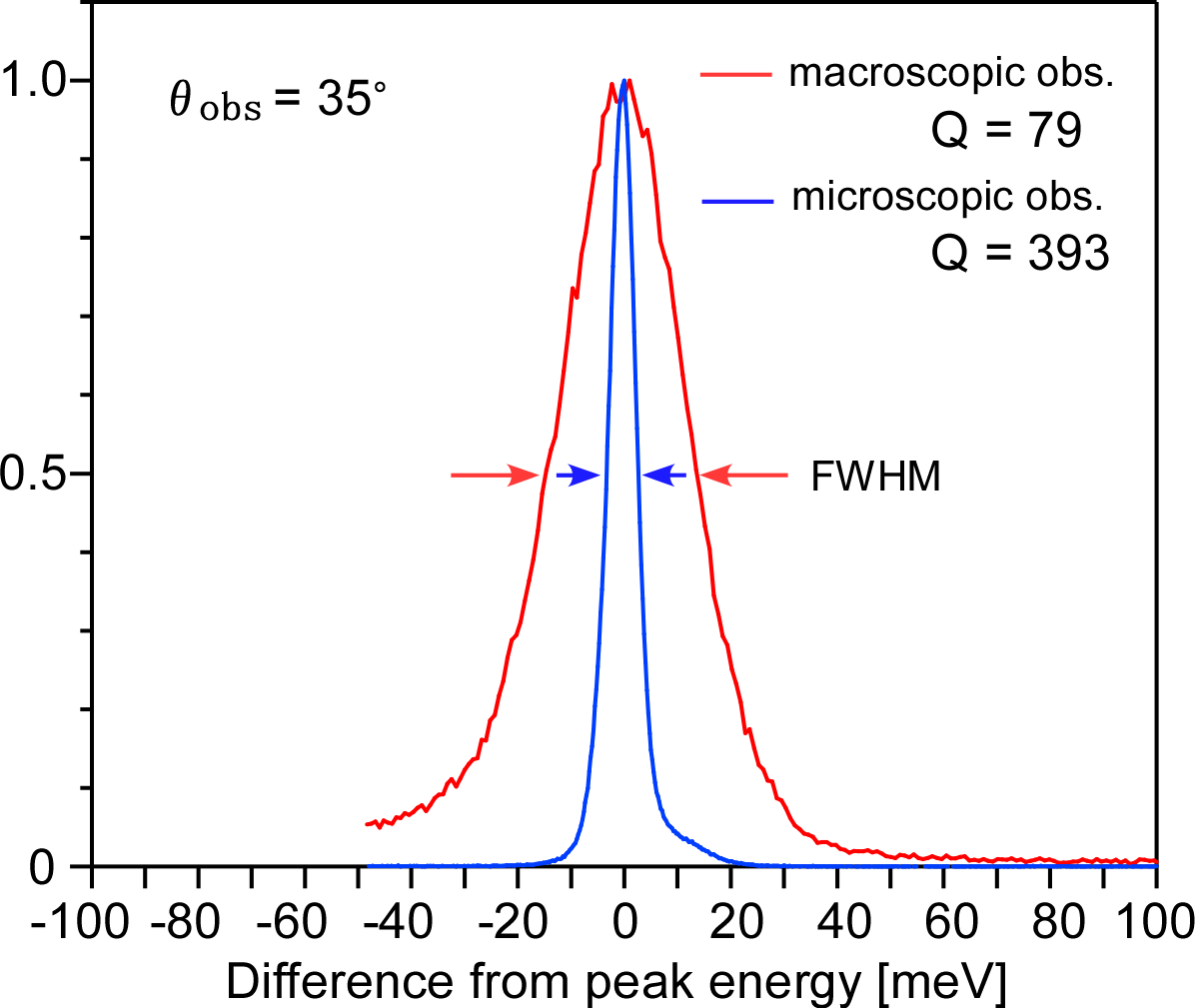}
  \caption{PL spectra of the NPL microcavity measured by the macroscopic observation with the spatial resolution of $\simeq$ 0.6 mm (red line) and that measured by the microscopic observation with the spatial resolution of $\simeq$ 30 $\mu$m (blue line). Both spectra were measured at $\theta_{\rm{obs}}$ = 35$^\circ$.} 
  \label{C1}
\end{figure}
\subsubsection{C2. $E$-$\theta_{\mathrm{obs}}$ map of the polariton populations}
Figure~S\ref{population} is the $E$-$\theta_{\mathrm{obs}}$ map of the polariton populations $D_\mathrm{\theta}$($E$). The scale bar is adjusted to show the change of the population distributions near 50$^\circ$. Although difficult to see, there is a yellow spot at the 50$^\circ$ at the energy of $E_{2}$. The yellow spot (medium population) is surrounded by a green area (slightly lower than the medium population). This suggests that the polariton populations at the spot are enhanced by the LO-phonon-assisted relaxation accompanied by two LO-phonon emissions.
 \begin{figure}[htb]
 \centering
 \includegraphics[width=2.9 in]
      {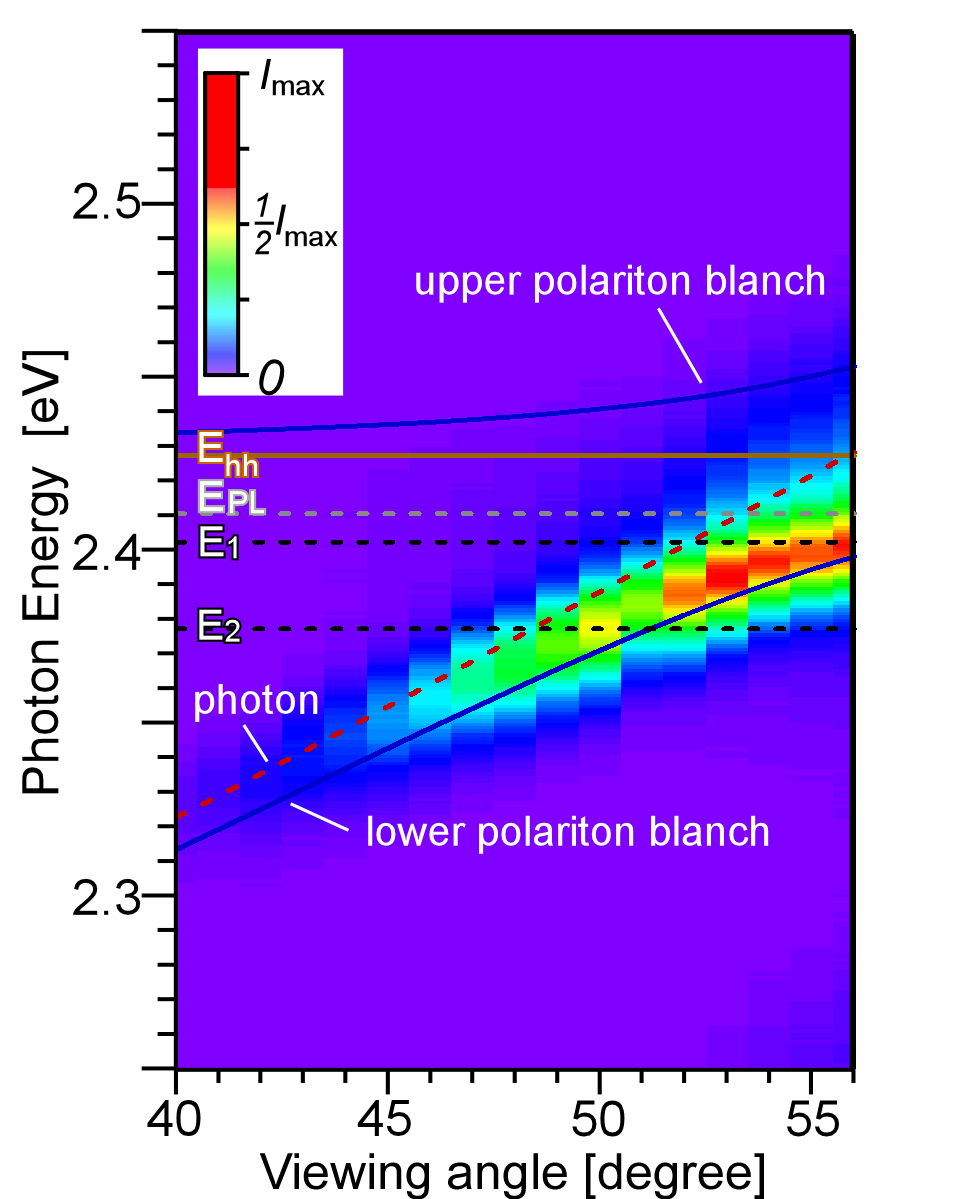}
 \caption{$E$-$\theta_{\mathrm{obs}}$ map of the polariton populations $D_\mathrm{\theta}$($E$).} 
 \label{population}
\end{figure}
\end{suppinfo}
\bibliography{achemso-demo}
\end{document}